\documentclass[12pt,reqno]{amsart}
\usepackage{amssymb}
\usepackage{amsfonts}
\usepackage{amssymb,latexsym}
\usepackage{enumerate}
\usepackage{mathrsfs}
\usepackage{url}
\usepackage{tikz}
\usetikzlibrary{calc, arrows.meta}

\makeatletter
\@namedef{subjclassname@2020}{%
  \textup{2020} Mathematics Subject Classification}
\makeatother

  [2002/01/19 v2.2g %
    AMS font definitions%
  ]
\DeclareFontFamily{U}{euf}{}
\DeclareFontShape{U}{euf}{m}{n}{%
  <5><6><7><8><9>gen*eufm%
  <10><10.95><12><14.4><17.28><20.74><24.88>eufm10%
  }{}
\DeclareFontShape{U}{euf}{b}{n}{%
  <5><6><7><8><9>gen*eufb%
  <10><10.95><12><14.4><17.28><20.74><24.88>eufb10%
  }{}

  [2002/01/19 v2.2g %
    AMS font definitions%
  ]
\DeclareFontFamily{U}{msb}{}
\DeclareFontShape{U}{msb}{m}{n}{%
  <5><6><7><8><9>gen*msbm%
  <10><10.95><12><14.4><17.28><20.74><24.88>msbm10%
  }{}

  [2002/01/19 v2.2g %
    AMS font definitions%
  ]
\DeclareFontFamily{U}{msa}{}
\DeclareFontShape{U}{msa}{m}{n}{%
  <5><6><7><8><9>gen*msam%
  <10><10.95><12><14.4><17.28><20.74><24.88>msam10%
  }{}

\theoremstyle{definition}

\numberwithin{equation}{section} 

\begin{document}

\title[On path integrals for wave functions taking $p$-adic values]
{On path integrals for wave functions taking $p$-adic values}



\author{Su Hu}
\address{Department of Mathematics, South China University of Technology, Guangzhou, Guangdong 510640, China}
\email{mahusu@scut.edu.cn}

\author{Min-Soo Kim}
\address{Department of Mathematics Education, Kyungnam University, Changwon, Gyeongnam 51767, Republic of Korea}
\email{mskim@kyungnam.ac.kr}




\begin{abstract}
In this paper, we construct a $p$-adic path integral via $p$-adic multiple integrals. This integral describes the evolution of a wave function $\Psi(x)$, defined as a map from a domain in $\mathbb{C}_p$ to $\mathbb{C}_p$. Unlike the standard $p$-adic quantum mechanics formulated for functions $\mathbb{Q}_p \to \mathbb{C}$, our construction works directly on $\mathbb{C}_p$-valued wave functions. Using $p$-adic Dirac delta measures and $p$-adic additive characters, we define the propagator as a finite-partition multiple integral, which allows explicit iterative integration. As an application, we compute the Feynman propagator for free particles and obtain a closed-form expression analogous to the classical counterpart. Our method provides a discrete, constructive alternative that complements the existing distribution-theoretic frameworks for $p$-adic functional integration.
\end{abstract}
 \maketitle

\section{Introduction}

\subsection{Background and motivation}

Throughout this paper we shall use the following notations.
\begin{equation*}
\begin{aligned}
\qquad \mathbb{N}  ~~~&- ~\textrm{the set of positive integers}.\\
\qquad \mathbb{C}  ~~~&- ~\textrm{the field of complex numbers}.\\
\qquad p  ~~~&- ~\textrm{an odd rational prime number}. \\
\qquad\mathbb{Z}_p  ~~~&- ~\textrm{the ring of $p$-adic integers}. \\
\qquad\mathbb{Q}_p~~~&- ~\textrm{the field of fractions of}~\mathbb Z_p.\\
\qquad\mathbb C_p ~~~&- ~\textrm{the completion of a fixed algebraic closure}~\overline{\mathbb Q}_p~ \textrm{of}~\mathbb Q_{p}.
\end{aligned}
\end{equation*}

In classical quantum mechanics, the evolution for the state of a particle $\Psi(\vec{r},t)=\Psi(x,y,z,t)$ is described by the Schr\"odinger equation
\begin{equation}\label{(1)}
i\hbar \frac{\partial}{\partial t} \Psi(\vec{r},t)=\hat{H}\Psi(\vec{r},t),
\end{equation}
where $\hat{H}=-\frac{{\hbar}^{2}}{2m} \Delta +V(\vec{r}\,)$ is the Hamiltonian, $\hbar$ is the Planck constant and
$$\Delta=\frac{{\partial}^2}{\partial x^2}+\frac{{\partial}^2}{\partial y^2}+\frac{{\partial}^2}{\partial z^2}$$
is the Laplacian. 
For $t\geq 0$, as interpreted by Born, $|\Psi(\vec{r},t)|^{2}$ is the probability density of location of a particle at a point $\vec{r}=(x,y,z)\in\mathbb{R}^{3}$.
So we have $$\iiint_{\mathbb{R}^{3}}|\Psi(x,y,z,t)|^{2}dxdydz=1,$$ which is equivalent to saying $\Psi(x,y,z,t)\in L^{2}(\mathbb{R}^{3})$.
In conclusion, the spacetime in classical quantum mechanics is described by real numbers. 

Different from the real numbers, the $p$-adic fields have many unusual properties both in geometry and analysis. For example, in $\mathbb{C}_{p}$, every triangle is isosceles and every point in an open disc is a center (see \cite[pp. 5--6]{Ko}). Furthermore, a series $\sum_{n=1}^{\infty} a_n$ converges if and only if $a_n \to 0$ as $n \to \infty$. These phenomena all arise from the non-archimedean property $|x+y|_p \leq \max\{|x|_p, |y|_p\}$.

It is precisely these exotic properties that motivate their physical application. In his article ``Number theory as the ultimate physical theory'' \cite{Volovich}, Volovich contended that at the Planck scale, the usual notion of spacetime breaks down and standard archimedean geometry becomes inapplicable. Since the Planck length represents a fundamental limit to measurable distances, the archimedean axiom is rendered physically unrealistic. Therefore, a non-archimedean geometry, such as that provided by the $p$-adic number field, must be considered. This allows for the construction of physical theories where the fundamental entities are numbers themselves, and where quantum fluctuations of the number field could lead to domains with $p$-adic or finite geometry, especially in regimes like gravitational collapse or the cosmological singularity.

In 1989, Vladimirov and Volovich \cite{Vladimirov} initiated the study of quantum mechanics over the $p$-adic rational field $\mathbb{Q}_p$. In this framework, every nonzero $p$-adic number $x \in \mathbb{Q}_p$ admits a unique canonical expansion of the form:
\begin{equation}\label{p-numb}
x = p^{\nu} \sum_{n=0}^\infty a_n p^n,
\end{equation}
where $\nu = \nu(x) \in \mathbb{Z}$, and the coefficients $a_n$ are integers satisfying $0 \leq a_n \leq p-1$ with $a_0 \neq 0$. Using this representation, the fractional part $\{x\}$ of $x \in \mathbb{Q}_p$ is defined as follows:
$$
\{x\} = 
\begin{cases}
0, & \text{if } \nu(x) \geq 0 \text{ or } x = 0, \\
p^\nu \left( a_0 + a_1 p + \cdots + a_{|\nu|-1} p^{|\nu|-1} \right), & \text{if } \nu(x) < 0.
\end{cases}
$$
Let $$\chi(x)=e^{2\pi i\{x\}},\quad x\in\mathbb{Q}_{p}$$ be the complex valued $p$-adic exponential, which is an additive character on $\mathbb{Q}_{p}$. 
Vladimirov and Volovich considered $p$-adic $L^2$-functions $\Psi(x)$ mapping from $\mathbb{Q}_p$ to $\mathbb{C}$. The evolution of a quantum state $\Psi(x)$ is described by a formal path integral:
\begin{equation}\label{1.1}
\Psi(y) = \int_{\mathbb{Q}_p} U(y,t;x,t_0) \Psi(x)  dx, \quad \Psi \in L^2(\mathbb{Q}_p).
\end{equation}
Here the kernel (propagator) $U(y,t;x,t_{0})$ is defined by  
\begin{equation}\label{1.2}
U(y,t;x,t_{0}) = \int \chi \left( \int_{t_{0}}^t L(\dot{x}(t), x(t))dt \right) \prod_{t_{0} \leq \tau < t} dx(\tau),
\end{equation}
where the integration is performed over classical trajectories with the boundary conditions $x(t_0) = x$, $x(t) = y$, and $L$ is a classical $p$-adic Lagrangian, $L(\dot{x}(t),x(t)) \in \mathbb{Q}_p$.
 
Let $B_n(a)$ be the disk with center at $a \in \mathbb{Q}_p$ and radius $p^n$:
\[
B_n(a) = \{ x \in \mathbb{Q}_p: |x - a|_p < p^n \}, \quad B_n(0) = B_n.
\]
In \cite{Zelenov}, Zelenov defined (\ref{1.2}) in a rigorous mathematical sense by introducing the $p$-adic line integral on a segment in $\mathbb{Q}_{p}$. Let $f: B_R \to \mathbb{C}_p$ be an analytic function on the disk $B_R$, with the power series expansion:
\begin{equation*}
f(x)=\sum_{n=0}^{\infty}a_{n}x^{n}, \quad a_{n}\in\mathbb{C}_{p}.
\end{equation*}
A primitive (or antiderivative) of $f$, denoted here by $F$, is defined via term-by-term integration of the series:
\begin{equation*}
F(x) = \sum_{n=0}^{\infty} \frac{a_n}{n+1} x^{n+1}.
\end{equation*}
This series converges on a disk $B'$ (which typically has the same radius $R$ as the original series). For any segment $[a, b] \subset B'$, the integral of $f$ over $[a, b]$ is then defined by (see \cite[(4.5)]{Zelenov}):
\[
\int_a^b f(x)  dx = F(b) - F(a).
\]

\subsection{Subdivision of a line segment}\label{Sec. 2.1}
In \cite[Sec. VI A]{Zelenov}, Zelenov introduced a method for subdividing a line segment in $\mathbb{Q}_p$. First, recall that a linear order can be defined on $\mathbb{Q}_p$ via a continuous injective map $\phi: \mathbb{Q}_p \to \mathbb{R}$ (see \cite[Lemma 3.1]{Zelenov}). Specifically, we define $a < b$ if and only if $\phi(a) < \phi(b)$, and accordingly define a segment as  
\[
[a,b] := \{ x \in \mathbb{Q}_p : a \leq x \leq b \}.
\]

Following Zelenov's construction, let $t \in \mathbb{Q}_p$ with $|t|_p = p^{-N-1}$ for some integer $N \ge 0$. Since $p^{-N-1} < p^N$, we have $t \in B_N$, where $B_N = \{x \in \mathbb{Q}_p : |x|_p < p^N\}$ is the open disk of radius $p^N$ centered at $0$. Moreover, by Zelenov's example, the open disk $B_N$ is contained in the segment $[0, -p^{-N}]$ in the linear order defined above, and the difference is only the endpoint $-p^{-N}$. Therefore, the segment $[0,t]$ is contained in $B_N$, and consequently, for every $x \in [0,t]$, we have $|x|_p < p^N$. According to \cite[Lemma 1.1]{Zelenov}, for any $n \leq N$, there exists a unique canonical covering of $B_N$ by disjoint disks of radius $p^n$:
\[
B_N = \bigcup_{a \in A} B_n(a), \quad \text{and} \quad B_n(a_k) \cap B_n(a_l) = \emptyset \ (k \neq l),
\]
where the index set $A$ has cardinality $\#A = p^{N-n}$.
If we write an element $a \in A$ as
\[
a = p^{-r}(a_0 + a_1 p + \cdots + a_{r-n-1} p^{\,r-n-1}),
\]
then the corresponding disk can be expressed in segment form:
\begin{equation}\label{2.2-1}
B_n(a) \text{ corresponds to the segment } [a,\, a-p^{-n}),
\end{equation}
with the understanding that the right endpoint is excluded when \(B_n(a)\) is viewed as an open disk.
The endpoints of these segments form a partition of the original segment $[0, -p^{-N}] \supset [0,t]$ into subsegments of length $-p^{-n}$. Enumerating these endpoints in increasing order gives
\begin{equation}\label{2.2-2}
0 = t_0 < t_1 < \cdots < t_k = -p^{-N}, \quad \text{where } k = p^{N-n}.
\end{equation}
Within this partition, there exists some $t_i$ such that
\[
t_{i-1} < t \leq t_i.
\]

The remainder of this paper is organized as follows. In Section II, we introduce the framework of wave functions taking values in $\mathbb{C}_p$, compare our approach with the earlier $p$-adic-valued quantum mechanics developed by Khrennikov, and construct the propagator via $p$-adic multiple integrals using Dirac delta measures. We also discuss the statistical interpretation of $\mathbb{C}_p$-valued wave functions in light of non-Archimedean probability theory. In Section III, we apply our construction to free particles, compute the propagator explicitly, and compare our finite-partition multiple-integral method with the distribution-theoretic Gaussian integration frameworks of Albeverio, Khrennikov and co-workers.

\section{Wave functions taking values in $\mathbb{C}_{p}$}

\subsection{Motivation and the non-Archimedean framework}
As noted in \cite[p. 66]{Vladimirov}, $p$-adic analysis can be developed in two distinct contexts: one considers functions from $\mathbb{Q}_{p}$ to $\mathbb{C}$, while the other considers functions from $\mathbb{C}_{p}$ to $\mathbb{C}_{p}$. In this work, differing from the approaches in \cite{Vladimirov} and \cite{Zelenov}, we consider the $p$-adic wave functions $\Psi(x)$ as a map from a domain in $\mathbb{C}_{p}$ to $\mathbb{C}_{p}$.

In this setting, a fundamental difficulty arises: there is no general theory of a ``Laplacian'' operator for $\mathbb{C}_{p}$-valued functions. Consequently, standard $p$-adic spectral theory lacks the tools to establish the Hermitian property of an operator in a general way, other than by explicitly constructing its eigenbasis (cf.~\cite{Vishik} or~\cite{Kochubei}). This implies that for a quantum system described within our framework, we cannot directly obtain the energy levels and eigenstates by solving differential equations like \eqref{(1)}. It is for this reason that we turn to a path integral formulation to describe the quantum dynamics.

It should be emphasized that the latter framework, quantum mechanics with $p$-adic valued wave functions, was already systematically developed by Khrennikov in a series of works, most notably in Ref.~\cite{Khrennikov1991} (see also Refs.~\cite{Khrennikov1990, Khrennikov1994book}). In that formalism, the underlying space is $\mathbb{Q}_p^n$, and the wave functions take values in a quadratic extension of \(\mathbb{Q}_p$ (or more generally in $\mathbb{C}_p$). Khrennikov introduced a non-Archimedean Hilbert space based on orthogonal Banach spaces, defined $p$-adic distributions via analytic functionals and Laplace transforms, and constructed Gaussian distributions that serve as integral kernels for pseudodifferential operators. In particular, the Schr\"odinger equation for $p$-adic valued functions governs the evolution of wave functions belonging to the Hilbert space $L_2(\mathbb{Q}_p^n, \nu)$ defined by a Gaussian measure $\nu$ on $\mathbb{Q}_p^n$, and the evolution was given by integral operators with $p$-adic symbols.

Our present approach differs from Khrennikov's in several essential aspects. First, we work directly on domains in $\mathbb{C}_p$ (not merely $\mathbb{Q}_p^n$) and do not rely on quadratic extensions or global Gaussian measures. Second, instead of defining the propagator via distribution-theoretic Laplace transforms and functional integration over infinite-dimensional spaces, we construct the path integral as a \textbf{finite-partition multiple integral} using $p$-adic Dirac delta measures (Eq.~\eqref{1.8}). The delta measures localize the intermediate positions, allowing us to perform iterative integrations explicitly and obtain a closed-form expression for the free propagator without invoking an infinite-dimensional Gaussian measure or a pseudo-differential symbol calculus. This discrete, stepwise construction is closer in spirit to the original Feynman--Kac idea, but adapted to the non-Archimedean setting where the usual limiting procedures (e.g., Riemann sums) are replaced by exact evaluations due to the ultrametric nature of the delta distributions. Our method thus provides an alternative route that complements the existing measure-theoretic and distribution-theoretic frameworks, and it yields explicit analytic formulae that are directly comparable to classical real-time path integrals.

\subsection{The propagator via $p$-adic multiple integrals}
Now let $\mu$ be a $p$-adic measure on $\mathbb{Z}_{p}$ (see \cite[p. 36, Definition]{Ko}). For any continuous function $f$ from $\mathbb{Q}_{p}$ to $\mathbb{C}_{p}$, $I_{\mu}(f)=\int_{\mathbb{Z}_{p}}f(x)d\mu(x)$, the $p$-adic integration with respect to $\mu$ is defined as a limit of the Riemann sums (see \cite[p. 39, Theorem 6]{Ko}). Suppose $\Psi(x)$ is locally analytic on $\mathbb{Z}_{p}$ (recall that a $p$-adic function $h: D\to\mathbb{C}_{p}$ is said to be locally analytic, if for each $a\in D$, there is a neighborhood $V\subset D$ of $a$ such that $h|_{V}$ is analytic (see \cite[p. 69, Definition 25.2]{SC}). Then the evolution of the state $\Psi(x)$ is defined by the integral
\begin{equation}\label{1.4}
\Psi(y)=\int_{\mathbb{Z}_{p}}U(y,t;x,t_{0})\Psi(x)d\mu(x),
\end{equation}
for a line segment $[t_{0}, t]\subset \mathbb{Q}_{p}$ with $x(t_0) = x$ and $x(t) = y$. In the following, we shall describe the kernel $U(y,t;x,t_{0})$.

Let $d$ be any point in $\mathbb{Q}_{p}$ and $U$ be any open subset of $\mathbb{Q}_{p}$. The $p$-adic Dirac delta measure is defined as follows:
\begin{equation}\label{1.5}
\delta_{d}(U)= \begin{cases}
1&\textrm{if}~d\in U,\\
0&\textrm{if}~d\notin U.
\end{cases}
\end{equation}
This measure cannot be constructed in $\mathbb{R}$ or $\mathbb{C}$ in the same way, because singletons are not open in the Archimedean topology. In $\mathbb{Q}_p$, every singleton $\{d\}$ is clopen, so the set function above is indeed a measure. Applying (\ref{1.5}), for any continuous function $f$ on $\mathbb{Q}_{p},$ we get  
\begin{equation}\label{1.6}
I_{\delta_{d}}(f)=\int_{\mathbb{Q}_{p}}f(x)d\delta_{d}(x)=f(d).
\end{equation}
It needs to mention that by applying the Dirac measure, we may obtain analytic continuations of Serre's $p$-adic family of Eisenstein series (see \cite[p. 2361]{Panchishkin} or \cite[p. 613]{HK2015}). In the present work, we extend the domain of definition of the delta measure from $\mathbb{Z}_p$ to the whole field $\mathbb{Q}_p$; this extension is harmless because the measure is purely atomic and the integration of a continuous function against it is still given by evaluation at the point $d$.

Now consider the propagator $U(y,t;x,t_{0})$ introduced in (\ref{1.4}).  
Let $E=\{x\in\mathbb{C}_{p}: |x|_{p}<p^{-\frac{1}{p-1}}\}$ be the region of convergence for the $p$-adic exponential function  
$$e_{p}(x)=\sum_{n=0}^{\infty}\frac{x^{n}}{n!}$$  
(see \cite[p. 70, Theorem 25.6]{SC}).  
For any nonzero $a\in\mathbb{C}_{p}$, the $p$-adic additive character with values in $\mathbb{C}_{p}$ is defined as  
$$\chi_{a}(x)=e_{p}(ax),$$  
which is well-defined for $x\in B^{(a)}:=\{y\in\mathbb{C}_{p}: |y|_{p}<|a|^{-1}_{p}\,p^{-\frac{1}{p-1}}\}$.  
If $a=0$, then $\chi_0(x)\equiv 1$ for all $x\in\mathbb{C}_{p}$, and the convergence condition is vacuous.

Let $V(x)$ be an analytic function from a domain in $\mathbb{C}_{p}$ to $\mathbb{C}_{p}$, and let $x(t)$ be an analytic curve mapping the line segment $[0, t]\subset \mathbb{Q}_{p}$ into $\mathbb{Z}_{p}$.  
Suppose $|t|_{p}=p^{-N-1}$ for some integer $N\ge 0$. Then $t\in B_N = \{x\in\mathbb{Q}_p: |x|_p < p^N\}$. Following Section \ref{Sec. 2.1}, we take the canonical covering of $B_N$ by disks of radius $p^n$ ($n\le N-1$), which yields a partition of $[0,t]$ into subsegments of $p$-adic length $-p^{-n}$; enumerating the endpoints in increasing order gives
\begin{equation}\label{2.2-2b}
0 = t_0 < t_1 < \cdots < t_k \quad \text{with } k = p^{N-n},
\end{equation}
where each subinterval $[t_i,t_{i+1}]$ has $p$-adic length $\Delta_n = t_{i+1}-t_i$ satisfying $|\Delta_n|_p = p^n$. We set
$$\epsilon_{n}=p^{n},\qquad x_{i}=x(t_{i})\;\; (0\leq i\leq k),$$
with $x=x_{0}$ and $y=x_k$.

For a particle of mass $m$ moving in a potential field $V(x)$, the Lagrangian is  
$$L(x(t), \dot{x}(t))=\frac{m}{2}\dot{x}^{2}(t)-V(x(t)).$$  
Using the absolute value of the time step as a scale, we define the discrete action over a subinterval as
\begin{equation}
\begin{aligned}
S[i, i+1] &:= \frac{m}{2} \frac{(x_{i+1} - x_{i})^2}{\epsilon_{n}} - V \left( \frac{x_i + x_{i+1}}{2} \right).
\end{aligned}
\end{equation}
We assume that the paths and the parameter \(a\in\mathbb{C}_p\) are chosen so that every discrete action term, after multiplication by \(a\), lies in the convergence domain of the exponential, i.e.,
\begin{equation}\label{conv-cond}
|a\,S[i,i+1]|_p < p^{-\frac{1}{p-1}} \quad \text{for all } i=0,\dots,k-1.
\end{equation}
This condition ensures that \(\chi_a(S[i,i+1])\) is well-defined for the chosen \(a\).

The $n$th propagator $U^{(n)}(y,t;x,t_0)$ is then defined by the $p$-adic multiple integral:
\begin{equation}\label{1.8}
\begin{aligned}
U^{(n)}(y,t;x,t_0) &= \int_{x}^{y} \chi_a \bigl( S[x(t)] \bigr) \, \mathcal{D}^{(n)}[x(t)] \\
&=\int_{\mathbb{Q}_p} \cdots \int_{\mathbb{Q}_p} 
   \chi_a \Biggl( \sum_{i=0}^{k-1} S[i, i+1] \Biggr) \\
&\quad \times d\delta_{\frac{x_{0}+x_{2}}{2}}(x_{1})\,d\delta_{\frac{1}{3}x_{0}+\frac{2}{3}x_{3}}(x_{2})\cdots d\delta_{\frac{1}{k}x_{0}+\frac{k-1}{k}x_{k}}(x_{k-1}).
\end{aligned}
\end{equation}
Here the intermediate points $x_1,\dots,x_{k-1}$ are integrated over the whole field $\mathbb{Q}_p$, whereas the boundary points $x_0=x$ and $x_k=y$ remain fixed in $\mathbb{Z}_p$. Since each Dirac delta measure is supported on a single point, the integrals are perfectly well defined on the non-compact space $\mathbb{Q}_p$, and the extension of the integration domain from $\mathbb{Z}_p$ to $\mathbb{Q}_p$ does not affect the evolution equation \eqref{1.4}, whose initial-position measure $\mu$ is still supported on $\mathbb{Z}_p$.

This construction is illustrated in the following figures:
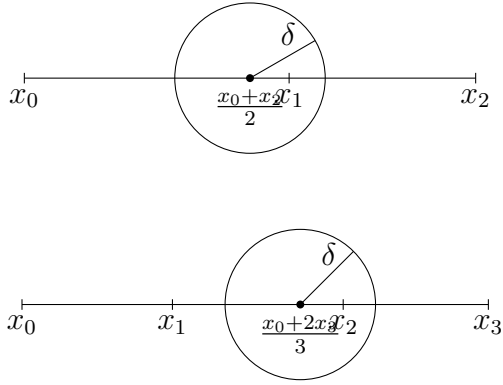
\begin{figure}[h!]
    \centering
    \begin{tikzpicture}
        \coordinate (x0) at (-3, 0);
        \coordinate (x2) at (3, 0);
        \coordinate (midpoint) at (0, 0);
        \coordinate (x1) at (0.52, 0);
        
        \draw[-] (x0) -- (x2);
        
        \draw ([yshift=-2pt]x0) -- ([yshift=2pt]x0);
        \node[below] at (x0) {$x_0$};
        
        \draw ([yshift=-2pt]x1) -- ([yshift=2pt]x1);
        \node[below] at (x1) {$x_1$};
        
        \draw ([yshift=-2pt]x2) -- ([yshift=2pt]x2);
        \node[below] at (x2) {$x_2$};
        
        \coordinate (center) at (midpoint);
        
        \fill (center) circle (1.5pt);
        \draw (center) circle (1cm);
        
        \draw[-] (center) -- ++(30:1cm);
        
        \node[above] at ($(center)+(30:0.59cm)$) {$\delta$};
        \node[below] at (center) {$\frac{x_0+x_2}{2}$};
    \end{tikzpicture}
    
    \vspace{1cm} 
    
    \begin{tikzpicture}
        \coordinate (x0) at (-3, 0);
        \coordinate (x3) at (3.2, 0);
        \coordinate (ratio_point) at (0.7, 0);
        \coordinate (x1) at (-1.0, 0);
        \coordinate (x2) at (1.27, 0);
        
        \draw[-] (x0) -- (x3);
        
        \draw ([yshift=-2pt]x0) -- ([yshift=2pt]x0);
        \node[below] at (x0) {$x_0$};
        
        \draw ([yshift=-2pt]x1) -- ([yshift=2pt]x1);
        \node[below] at (x1) {$x_1$};
        
        \draw ([yshift=-2pt]x2) -- ([yshift=2pt]x2);
        \node[below] at (x2) {$x_2$};
        
        \draw ([yshift=-2pt]x3) -- ([yshift=2pt]x3);
        \node[below] at (x3) {$x_3$};
        
        \coordinate (center) at (ratio_point);
        
        \fill (center) circle (1.5pt);
        \draw (center) circle (1cm);
        
        \draw[-] (center) -- ++(45:1cm);
        
        \node[above] at ($(center)+(45:0.54cm)$) {$\delta$};
        \node[below] at (center) {$\frac{x_0+2x_3}{3}$};
    \end{tikzpicture}
    
    \caption{The process of $p$-adic path integral}
    \label{fig:x0x1x2x3_circle}
\end{figure}

\subsection{Remarks on probability interpretation}
Since our wave functions take values in \(\mathbb{C}_p\) rather than in \(\mathbb{C}\), the standard Born probability interpretation, which relies on the real-valued quantity \(|\Psi|^2\), is not directly applicable. This raises the fundamental question of how to assign a statistical meaning to the \(p\)-adic-valued quantum states. A systematic framework for addressing this issue was developed by Khrennikov in the context of non-Archimedean probability theory \cite[Chapter VI]{Khrennikov1994book}. In that approach, the usual Kolmogorov axiomatics based on real-valued measures is replaced by a more general frequency theory of probability, where relative frequencies may stabilize with respect to a \(p\)-adic topology rather than the usual real topology. In this framework, probabilities can assume values in \(\mathbb{Q}_p\) (or its quadratic extensions), and the standard real interval \([0,1]\) is replaced by the entire field \(\mathbb{Q}_p\) (since the closure of rational frequencies in the unit interval with respect to the \(p\)-adic metric is the whole \(\mathbb{Q}_p\)). Thus, the coefficients \(|\Psi(x)|^2\) or the expansion coefficients of a state in a given basis can be interpreted as \(p\)-adic probabilities of observing certain outcomes, provided that the corresponding statistical experiments exhibit \(p\)-adic stabilization of relative frequencies. This interpretation has been applied to both quantum mechanics and quantum field theory with \(p\)-adic-valued wave functions, as discussed in Refs.~\cite{Khrennikov1991, Khrennikov1994book} and subsequent works. In particular, Khrennikov showed that, for such states, relative frequencies may stabilize in the \(p\)-adic topology even when they fail to stabilize in the real topology, leading to a \(p\)-adic probability interpretation of the expansion coefficients (see \cite[Chapter VII]{Khrennikov1994book}).  We adopt this viewpoint in the present work. Although we do not directly compute probabilities, our path integral construction provides a propagator whose matrix elements, when squared in the \(p\)-adic norm (or via the quadratic extension), can be interpreted as transition probabilities in the non-Archimedean probability sense.  A detailed physical interpretation of specific models will be addressed in future investigations.

\section{Free particles}\label{free}

\subsection{Explicit computation of the propagator}
In this section, we shall calculate the propagator for free particles, and we will see that the result is very similar to the classical situations (see \cite[Sec. 8.4]{Shankar}).

For free particles with mass $m$, the potential function $V(x)\equiv 0$ for $x\in\mathbb{Q}_{p}$. Thus the local convergence condition \eqref{conv-cond} becomes
\[
\left| a\,\frac{m}{2}\frac{(x_{i+1}-x_i)^2}{\epsilon_n}\right|_p < p^{-\frac{1}{p-1}} \quad \text{for all } i.
\]
Then the $n$th propagator $U^{(n)}(y,t;x,t_{0})$ from (\ref{1.8}) for free particles is
\begin{equation}\label{1.9b}
\begin{aligned}
U^{(n)}(y,t;x,t_{0})&=\int_{\mathbb{Q}_{p}}\int_{\mathbb{Q}_{p}}\cdots\int_{\mathbb{Q}_{p}}e_{p}\left(a\frac{m}{2}\sum_{i=0}^{k-1}\frac{(x_{i+1}-x_{i})^{2}}{\epsilon_{n}}\right)\\
&\quad \times d\delta_{\frac{x_{0}+x_{2}}{2}}(x_{1})\,d\delta_{\frac{1}{3}x_{0}+\frac{2}{3}x_{3}}(x_{2})\cdots d\delta_{\frac{1}{k}x_{0}+\frac{k-1}{k}x_{k}}(x_{k-1}).
\end{aligned}
\end{equation}

Next, we calculate the above integral. First, by (\ref{1.6}) we have
\begin{equation}
\begin{aligned}
&\quad\int_{\mathbb{Q}_{p}}e_{p}\left(a\frac{m}{2}\frac{(x_1-x_0)^{2}+(x_2-x_1)^{2}}{\epsilon_{n}}\right) d\delta_{\frac{x_{0}+x_{2}}{2}}(x_{1})\\
&=\int_{\mathbb{Q}_{p}}e_{p}\left(a\frac{m}{2\epsilon_{n}}\left[2\left(x_{1}-\frac{1}{2}(x_{0}+x_{2})\right)^{2}+\frac{1}{2}(x_{2}-x_{0})^{2}\right]\right) d\delta_{\frac{x_{0}+x_{2}}{2}}(x_{1})\\
&=e_{p}\left(a\frac{m}{2}\frac{1}{2\epsilon_{n}}(x_{2}-x_{0})^{2}\right).
\end{aligned}
\end{equation}

Similarly,
\begin{equation}
\begin{aligned}
&\quad\int_{\mathbb{Q}_{p}}\int_{\mathbb{Q}_{p}}e_{p}\left(a\frac{m}{2}\frac{(x_{1}-x_{0})^{2}+(x_2-x_1)^{2}+(x_{3}-x_{2})^{2}}{\epsilon_{n}}\right)\\
&\qquad \times d\delta_{\frac{x_{0}+x_{2}}{2}}(x_{1})\,d\delta_{\frac{1}{3}x_{0}+\frac{2}{3}x_{3}}(x_{2})\\
&=\int_{\mathbb{Q}_{p}}e_{p}\left(a\frac{m}{2\epsilon_{n}}\left[\frac{1}{2} (x_{2}-x_0)^{2}+(x_3-x_{2})^{2}\right]\right)d\delta_{\frac{1}{3}x_{0}+\frac{2}{3}x_{3}}(x_{2})\\
&=\int_{\mathbb{Q}_{p}}e_{p}\left(a\frac{m}{2\epsilon_{n}}\left[\frac{3}{2} \left(x_{2}-\frac{1}{3}x_{0}-\frac{2}{3}x_{3}\right)^{2}+\frac{1}{3}(x_3-x_{0})^{2}\right]\right)\\
&\quad\times d\delta_{\frac{1}{3}x_{0}+\frac{2}{3}x_{3}}(x_{2})\\
&=e_{p}\left(a\frac{m}{2}\frac{1}{3\epsilon_{n}}(x_3-x_{0})^{2}\right).
\end{aligned}
\end{equation}

Inductively, at the $(k-1)$-th step, we obtain
\begin{equation}\label{final_propagator}
\begin{aligned}
&\quad\int_{\mathbb{Q}_{p}}e_{p}\left(a\frac{m}{2\epsilon_{n}}\left[\frac{1}{k-1} (x_{k-1}-x_0)^{2}+(x_{k}-x_{k-1})^{2}\right]\right)\\
&\qquad \times d\delta_{\frac{1}{k}x_{0}+\frac{k-1}{k}x_{k}}(x_{k-1})\\
&=\int_{\mathbb{Q}_{p}}e_{p}\left(a\frac{m}{2\epsilon_{n}}\left[\frac{k}{k-1} \left(x_{k-1}-\frac{1}{k}x_{0}-\frac{k-1}{k}x_{k}\right)^{2}+\frac{1}{k}(x_k-x_{0})^{2}\right]\right)\\
&\quad\times d\delta_{\frac{1}{k}x_{0}+\frac{k-1}{k}x_{k}}(x_{k-1})\\
&=e_{p}\left(a\frac{m}{2}\frac{1}{k\epsilon_{n}}(x_k-x_{0})^{2}\right)=e_{p}\left(a\frac{m}{2}\frac{(x_{k}-x)^{2}}{p^{N}}\right),
\end{aligned}
\end{equation}
since $k\epsilon_{n}=p^{N-n}\cdot p^{n}=p^{N}$. 

The factor \(p^N\) arises naturally from the product \(k\epsilon_n\), which is independent of the subdivision fineness \(n\). Because \(|t|_p = p^{-N-1}\), we have \(p^N = p^{-1}|t|_p^{-1}\), and the propagator takes the form
\[
U^{(n)}(y,t;x,t_0) = e_p\left( a\frac{m}{2} \frac{(y-x)^2}{p^{-1}|t|_p^{-1}} \right)
= e_p\left( a\frac{m}{2} p |t|_p (y-x)^2 \right),
\]
which is the $p$-adic analogue of the classical free-particle propagator $$\exp\left(i m (y-x)^2 / (2t)\right)$$ if we identify \(p|t|_p\) with the inverse time scale.

\subsection{Comparison with distribution-theoretic frameworks}
Our finite-partition multiple-integral construction (Eq.~\eqref{1.8}) differs substantially from previously established $p$-adic functional integration and distribution-theoretic approaches. In essence, the Albeverio--Khrennikov framework is operator-theoretic, defining evolution via pseudo-differential operators and Gaussian measures, whereas ours is path-integral, computing propagators directly via discrete sums over paths. To see this in detail, note that Albeverio, Khrennikov and co-workers developed an infinite-dimensional $p$-adic Gaussian integration theory inspired by the Albeverio--Hoegh--Krohn idea of defining Feynman path integrals via Fourier transforms of bounded measures \cite{Albeverio1995}. In that framework, the $p$-adic Gaussian distribution on $\mathbb{Q}_p^\infty$ (or on nested Hilbert spaces $S_{-\infty}$) is introduced as a generalized function whose Laplace transform is an exponential quadratic form \cite{Albeverio1995}. The corresponding space $L_2(S_{-\infty}, d\nu)$ of square-integrable functions with respect to this Gaussian distribution is constructed via the orthogonal basis of Hermite polynomials, and the position and momentum operators are shown to be bounded \cite{Albeverio1996}. This approach is distribution-theoretic in nature: it relies on the isomorphism between distributions and analytic functions via the $p$-adic Laplace transform, and it defines the Gaussian integral as the action of a distribution on a test function, without any limiting procedure over partitions.

Earlier, Khrennikov also developed a theory of generalized functions on $p$-adic superspaces, where Gaussian ``supermeasures'' are defined using the same Parseval-equality technique \cite{Khrennikov1992}. These works provide powerful machinery for quantum field theory over $p$-adic numbers, but they all operate on the base field $\mathbb{Q}_p$ (or its quadratic extensions) and employ global analytic functionals and Laplace transforms to define integrals.

In contrast, our present construction works directly on domains in $\mathbb{C}_p$, avoids quadratic extensions, and does not require an infinite-dimensional Gaussian measure or a Laplace-transform formalism. Instead, we define the propagator as a \textbf{finite-partition multiple integral} with $p$-adic Dirac delta measures (Eq.~\eqref{1.8}). These delta measures localize the intermediate path points, allowing us to perform the integrations \textbf{iteratively and explicitly}, as demonstrated in the free-particle case (Eqs.~\eqref{1.9b} and~\eqref{final_propagator}), where the final expression reduces to a simple closed form involving only the endpoints. This stepwise discrete construction is closer to the original Feynman--Kac idea of path integration, but adapted to the non-Archimedean setting, where the ultrametric nature of the delta distributions replaces the usual limiting procedures by exact evaluations. Our method is thus complementary to the distribution-theoretic frameworks: it offers a direct, constructive alternative that avoids infinite-dimensional measure theory and symbol calculus, while producing analytic formulae comparable to those of classical real-time path integrals.

We emphasize that the boundedness of operators in the distribution-based $L_2$-spaces (as shown in \cite{Albeverio1996}) is a different aspect of the theory; our focus is on the \textbf{propagator} itself, not on operator spectra. The explicit free propagator we obtain in Eq.~\eqref{final_propagator} mirrors the classical form, and our iterative delta-integration technique may offer a new route to handle interactions in $\mathbb{C}_p$-valued quantum mechanics.

\section*{Author Declarations}
\textbf{Conflict of Interest}  The authors have no conflicts to disclose.

\section*{Data Availability}
Data sharing is not applicable to this article as no new data were created or analyzed in this study.

\section*{Acknowledgements} 
The authors are deeply grateful to the anonymous referee for the careful reading of this paper and for the many insightful suggestions. 

Su Hu is supported by the Natural Science Foundation of Guangdong Province, China (No. 2026A1515012852 and No. 2024A1515012337).  Min-Soo Kim is supported by the National Research Foundation of Korea (NRF) grant funded by the Korea government (MSIT) (No. RS-2025-23323090).


\begin{thebibliography}{00}

\bibitem{Albeverio1996} S. Albeverio and A.\,Yu. Khrennikov,
\textit{$p$-adic Hilbert space representation of quantum systems with an infinite number
of degrees of freedom},
Memorial issue for H. Umezawa.
Internat. J. Modern Phys. B \textbf{10} (1996), no. 13-14, 1665--1673.
	
\bibitem{Donoghue} J. Donoghue and L. Sorbo, 
\textit{A prelude to quantum field theory},
Princeton University Press, Princeton, NJ, 2022.
 
\bibitem{HK2015} S. Hu and M.-S. Kim, 
\textit{On $p$-adic analogue of Weil’s elliptic functions according to Eisenstein},
J. Number Theory \textbf{147} (2015), 605--619.

 
\bibitem{Ko} N. Koblitz,
\textit{$p$-adic Numbers, $p$-adic Analysis and Zeta-Functions}, 
2nd ed., Springer-Verlag, New York, 1984.

\bibitem{Kochubei} A.N. Kochubei, 
\textit{Harmonic oscillator in characteristic $p$},
Lett. Math. Phys. \textbf{45} (1998), no. 1, 11--20.


\bibitem{Khrennikov1990} A.\,Yu. Khrennikov,
\textit{Quantum mechanics over non-Archimedean number fields},
Teoret. Mat. Fiz. \textbf{83} (1990), no. 3, 406--418; translation in Theoret. and Math. Phys. \textbf{83}
(1990), no. 3, 623--632.

\bibitem{Khrennikov1991} A.\,Yu. Khrennikov,
\textit{$p$-adic quantum mechanics with $p$-adic valued functions},
J. Math. Phys. \textbf{32} (1991), no. 4, 932--937.

\bibitem{Khrennikov1992} A.\,Yu. Khrennikov,
\textit{Analysis on the $p$-adic superspace. I. Generalized functions: Gaussian
distribution},
J. Math. Phys. \textbf{33} (1992), no. 5, 1636--1642.

\bibitem{Khrennikov1994book} A.\,Yu. Khrennikov,
\textit{$p$-adic valued distributions in mathematical physics},
Mathematics and its Applications, 309.
Kluwer Academic Publishers Group, Dordrecht, 1994. 

\bibitem{Albeverio1995} A.\,Yu. Khrennikov,
\textit{Albeverio and Hoegh-Krohn approach to a $p$-adic functional integration},
J. Phys. A \textbf{28} (1995), no. 9, 2627--2635.

\bibitem{Panchishkin} A.A. Panchishkin, 
\textit{On $p$-adic integration in spaces of modular forms and its applications}, 
J. Math. Sci. (N. Y.) \textbf{115} (3) (2003) 2357--2377.

\bibitem{SC} W.H. Schikhof, 
\textit{Ultrametric Calculus: An Introduction to $p$-Adic Analysis}, 
Cambridge University Press, London, 1984.

\bibitem{Shankar} R. Shankar, 
\textit{Principles of quantum mechanics},
Second edition, Corrected reprint of the second (1994) edition,
Springer, New York, 2008. 

\bibitem{Vishik} M.M. Vishik, 
\textit{Non-archimedean spectral theory},
J. Sov. Math. \textbf{30} (1985), 2513--2554.

\bibitem{Vladimirov} V.S. Vladimirov and I.V. Volovich, 
\textit{$p$-adic quantum mechanics},
Comm. Math. Phys. \textbf{123} (1989), no. 4, 659--676.

\bibitem{Volovich} I.V. Volovich,
\textit{Number theory as the ultimate physical theory},
$p$-Adic Numbers Ultrametric Anal. Appl. \textbf{2} (2010), no. 1, 77--87.

\bibitem{Zelenov} E.I.  Zelenov,
\textit{$p$-adic path integrals},
J. Math. Phys. \textbf{32} (1991), no. 1, 147--152.


\end{thebibliography}
\end{document}